\documentclass[metabolites,article,accept,moreauthors,pdftex,12pt,a4paper]{mdpi}
\usepackage{epsfig,bm}
\usepackage{graphicx}
\usepackage{amsmath,amssymb,psfrag,textcomp}
\usepackage{amsthm}

\newcommand{\be}{\begin{equation}}
\newcommand{\ee}{\end{equation}}
\newcommand{\bd}{\begin{displaymath}}
\newcommand{\ed}{\end{displaymath}}

\newcommand{\beeq}[1] {\begin{equation}\begin{split}#1\end{split}\end{equation}}

\newcommand{\beaeq}[1] {\begin{eqnarray}#1\end{eqnarray}}

\lastpage{x}
\doinum{10.3390/------}
\pubvolume{xx}
\pubyear{2013}
\history{}
\usepackage{soul}
\Title{A Novel Methodology to Estimate Metabolic Flux Distributions in Constraint-Based Models
}

\Author{Francesco Alessandro Massucci $^{1,}$*, Francesc Font-Clos $^{2,3}$, Andrea De Martino $^{4,5,6}$ \mbox{and Isaac P\'erez Castillo $^{7}$}}

\address{%
$^{1}$ Department d'Enginyeria Qu\'\i mica, Universitat Rovira i Virgili, Tarragona 43007, Spain\\
$^{2}$ Centre de Recerca Matem\`atica, Edifici C, Campus Bellaterra, Bellaterra (Barcelona) E-08193, Spain; E-Mail: fontclos@crm.cat\\
$^{3}$ Department de Matem\`atiques, Universitat Aut\`onoma de Barcelona, Edifici C, Bellaterra~(Barcelona)~E-08193, Spain\\
$^{4}$ Dipartimento di Fisica, Sapienza Universit\`a di Roma, p. A. Moro 2, Rome 00185, Italy; \linebreak E-Mail: andrea.demartino@roma1.infn.it\\
$^{5}$ CNR/IPCF, UOS di Roma, Roma 00185, Italy\\
$^{6}$ Center for Life Nano Science@Sapienza, Istituto Italiano di Tecnologia, v. Regina Elena 291, \linebreak Rome 00161, Italy\\
$^{7}$ Department of Mathematics, King's College London, London WC2R 2LS, UK; \linebreak E-Mail: isaac.perez\_castillo@kcl.ac.uk}

\corres{E-Mail: francesco.massucci@urv.cat; \linebreak Tel.: +34-977-558-580; Fax +34-977-559-621.}

\abstract{
Quite generally, constraint-based metabolic flux analysis describes the space of viable flux configurations for a metabolic network as a high-dimensional polytope defined by the linear constraints that enforce the balancing of production and consumption fluxes for each chemical species in the system. In some cases, the complexity of the solution space can be reduced by performing an additional optimization, while in other cases, knowing the range of variability of fluxes over the polytope provides a sufficient characterization of the allowed configurations. There are cases, however, in which the thorough information encoded in the individual distributions of viable fluxes over the polytope is required. Obtaining such distributions is known to be a highly challenging computational task when the dimensionality of the polytope is sufficiently large, and the problem of developing cost-effective {\it ad hoc} algorithms has recently seen a major surge of interest. Here, we propose a method that allows us to perform the required computation heuristically in a time scaling {\it linearly} with the number of reactions in the network, overcoming some limitations of similar techniques employed in recent years. As a case study, we apply it to the analysis of the human red blood cell metabolic network, whose solution space can be sampled by different exact techniques, like Hit-and-Run Monte Carlo (scaling roughly like the third power of the system size). Remarkably accurate estimates for the true distributions of viable reaction fluxes are obtained, suggesting that, although further improvements are desirable, our method enhances our ability to analyze the space of allowed configurations for large biochemical reaction~networks.
}


\begin{document}

%%%%%%%%%%%%%%%%%%%%%%%%%%%%%%%%%%%%%%%%%%
\vspace{-12pt}
\section{Introduction}

The development of high throughput techniques now makes available a considerable number of high quality reconstructions of the metabolism of a variety of organisms, which include the stoichiometry of the biochemical reactions in the network and the underlying enzyme-gene associations \cite{Bowman1997, Feist2007, Thiele2010, Thiele2013}. Ideally, this information may be employed for kinetic modeling approaches that could shed light on issues, like the organization of a cell's metabolic phenotype and its robustness to perturbations (internal or exogenous) in a fully dynamical setting. Indeed, consider a metabolic network of $N$ coupled chemical reactions transforming $M$ metabolites and let $\bm{\xi}=\{\xi_{i}^{\mu}\}$ ($i=1,\ldots, N$; $\mu=1,\ldots, M$) denote the stoichiometric coefficients,
with the standard sign convention to distinguish substrates ($\xi_i^\mu<0$) from products \linebreak($\xi_i^\mu>0$) in each reaction, $i$. If one denotes by $\gamma^\mu$ the rate of change of the intracellular level of species $\mu$, due to exchanges between the cell's interior and the environment, then, under mass action kinetics, the intracellular concentration, $c^\mu$, of metabolite $\mu$ obeys the equation:
\beeq{
\frac{d c^\mu}{dt}=\sum_{i=1}^N\xi_i^\mu x^i-\gamma^\mu %\,,\quad \mu=1,\ldots, M
\label{eq:massbalan}
}
where $\gamma^\mu>0$ (resp.%please define -- defined in paragraph above
 $\gamma^\mu<0$) if there is a net out-take (resp. in-take) of species $\mu$.

Unluckily, addressing the above system in full generality requires knowledge about reaction mechanisms and kinetic constants (which specify how rates depend on concentrations), which is at best only partially available. (Besides, it is not entirely clear to us that, were that information fully at our disposal, simulating (\ref{eq:massbalan}) for a genome-scale reconstruction involving thousands of reactions and metabolites would be a sensible thing to do).

Computational studies of metabolic networks therefore generally assume that the cell operates at non-equilibrium steady-state (NESS) conditions, where the concentration of the metabolites is constant~\cite{Kauffman2003,Orth2010}, with the rationale that as long as one is interested in the behaviour for time scales shorter than genetic ones (minutes), the much faster equilibrating biochemistry can be assumed to be ``frozen'' in an NESS. Under this assumption, computing NESS fluxes more modestly requires the prescription of bounds for flux variability, \emph{i.e}., $x^i\in[m^{i},M^i]$ (which also encode for reaction reversibility assumptions), and for exchange rates, \emph{i.e}., $\gamma^\mu\in[m^\mu,M^\mu]$, with which conditions \eqref{eq:massbalan} can be written as:
\beaeq{
&&m^\mu\leq\sum_{i=1}^N\xi_{i}^\mu x^i\leq M^\mu\,,\quad\quad \mu=1,\ldots,M\label{eq:massbalance}\\
&&m^{i}\leq x^i\leq M^{i}\,,\quad\quad\quad\quad~~~~ i=1,\ldots,N
}
This set of inequalities is fairly general, since it may include metabolites involved only in internal reactions (for which $\gamma^\mu=0$), as well as exchanged species (\emph{i.e}., with $\gamma^\mu\neq0$).
%Henceforth, we will denote by $\mathcal{I}$ the set of metabolites with $\gamma^\mu=0$.
The solution space of Equation~(\ref{eq:massbalance}) is, in turn, given by:
%Thus given a metabolic reconstruction, that is, given the stoichiometry $\bm{\xi}$, the goal is to find a way to extract information on the set $S$ of all possible solutions for the vector of reaction rates $\bm{x}=(x^1,\ldots,x^N)$, \textit{viz.}
\beeq{
S=\left\{\bm{x}\in\mathbb{R}^N\text{ s.t. }m^\mu\leq\sum_{i=1}^N\xi_{i}^\mu x^i\leq M^\mu\,~\, (\mu=1,\ldots,M) \text{ and } m^{i}\leq x^i\leq M^{i}\,~\, (i=1,\ldots,N)\right\}
\label{eq:S}
}

The problem we address here concerns computational methods to sample $S$, in order to retrieve information on quantities like the distribution of the allowed values of each flux over the entire solution space. While most studies aiming at developing predictive power on metabolic phenotypes (especially for microbes) have been thus far based on coupling (\ref{eq:massbalance}) with a phenotypic optimization principle (the solving of which requires no sampling of $S$), it is being increasingly recognized that the information entailed by the structure of $S$ might provide key insight into different aspects of metabolic network analysis, from flux-flux correlations, to robustness, to perturbations, to (more subtly) optimal experiment design \cite{Schellenberger2009}. The task, however, presents many challenges, especially from the viewpoint of CPU costs. The running time of the paradigmatic algorithm to sample high-dimensional polytopes, such as $S$, namely Hit-and-Run Monte Carlo, is known to scale, in the best case, as the third power of the number of variables in the system \cite{Lovasz1999}, making it impractical for large enough networks. On the other hand, it has recently been shown that the heuristics of message-passing (MP) algorithms may provide a powerful alternative \cite{Braunstein2008}. In brief, MP-based protocols are designed to compute solutions to statistical inference problems, like estimating marginals of random variables, exploiting peculiar topological properties of the underlying graphs that describe the interdependence of variables (reactions in our case) and constraints (metabolites in our case). Indeed, when such a graph is locally tree-like, \emph{i.e}., it lacks short loops, statistical inference can be performed accurately by MP in linear time with the number of variables~\cite{Mezard2009}. This property has been used in \cite{Braunstein2008} to devise a highly efficient MP method to sample sets like $S$. Yet more recently \cite{Font2012}, a second MP algorithm has been proposed to overcome some of the limitations of the method of \cite{Braunstein2008}, most notably, the inapplicability of the latter to real valued stoichiometry and, perhaps more importantly, the accuracy problems that may arise when (some) fluxes are allowed to span several orders of magnitude.

Here, we build on the work presented in \cite{Font2012} and apply a MP methodology, which we call {\it weighted Belief Propagation} (wBP), to the analysis of the solution space of metabolic reaction networks. In particular, we focus on the metabolism of the human red blood cell (hRBC), a major benchmark for sampling tools, as its size ($N=46$, $M=34$) makes it possible to characterize its solution space $S$ by various methods and to compare the results. Specifically, we have evaluated results retrieved by wBP against those obtained by Hit-and-Run Monte Carlo to sample the polytope $S$ for the hRBC ({A note of caution here is needed regarding the words \textit{Monte Carlo}, which are used throughout the text, as well as in some of the references in a somewhat unspecific way. As is well-known, Monte Carlo techniques generically rely on repeated random sampling. Thus, Hit-and-Run is a Monte Carlo algorithm. The method used in \cite{Price2004} to sample a volume is also a Monte Carlo algorithm, albeit a different one, based on a rejection method. Methods to perform statistical inference by directly computing high-dimensional integrals, or sums, are often also Monte Carlo methods, as long as the integrals involved are evaluated by Monte Carlo integration. Therefore, by itself, \textit{Monte Carlo} is perhaps too generic a term and does not reveal the specifics of an algorithm. The reader is advised to check the references carefully when comparing the various techniques).
%It is well-known that, even though Hit-and-Run samples $S$ uniformly, its mixing time goes as $N^3$.
Because of the importance of Hit-and-Run as a mathematically controlled procedure with broad applicability, we have tried to devise a Monte Carlo method with computing times reduced as much as possible (the limit of standard Hit-and-Run techniques lying in their mixing time, known to scale cubically with the number of variables). The modified Hit-and-Run algorithm we have devised, based on a projection method, appears to be indeed optimized from this viewpoint with respect to, e.g., what was done in \cite{Almaas2004}. %This modification allows us to work directly in the subspace defined by the set of metabolites involved only in internal reactions.
We call the resulting method the \textit{Kernel Hit-and-Run} algorithm (KHR). As we will see, results obtained by wBP and KHR are in remarkable agreement among themselves (and with previous studies of the same problem by other, more costly, computational methods).

%%%%%%%%%%%%%%%%%%%%%%%%%%%%%%%%%%%%%%%%%%
\section{Methodology}
\vspace{-12pt}
\subsection{Mathematical Statement of the Problem}

Suppose that we are interested in estimating the probability distribution functions (PDFs) $P_i(x)$ for each reaction flux $i=1,\ldots,N$. By definition, they are given by:
\beeq{
P_{i}(x)=\frac{\text{Vol}(S_i(x))}{\text{Vol}(S)},\qquad S_i(x)=\{\bm{x}\in S\text{ s.t. } x^i=x\}
}
where $\text{Vol}(S)$ is the volume of set $S$. $P_i(x)$ can be mathematically written as an integral over all fluxes, but $x_i$, of a set of functions enforcing the constraints that define the polytope, namely (\ref{eq:massbalan}). Denoting by $F_\mu$ these functions ($\mu=1,\ldots, M$), we can write $P_i(x)$ for flux $i$ as:
\beeq{
P_i(x)=\frac{1}{Z_i}\int_{D_{\backslash i}}d\bm{x}_{\backslash i}\prod_{\mu=1}^M F_{\mu}\left(\sum_{\ell=1}^N\xi_\ell^\mu x^\ell\right)\,
\label{eq:pdfs}
}
where $\bm{x}_{\backslash i}=(x^1,\ldots, x^{i-1},x^{i+1},\ldots, x^N)$ denotes the collective integration variable, $D_{\backslash i}=\times_{\ell(\neq i)}^N[m^\ell,M^\ell]$ is the domain of integration (the set-product of the ranges of variability of all fluxes, except flux $i$) and $Z_i\equiv\int dx P_i(x)$ is a normalisation constant, so that each $P_i(x)$ is properly normalised to one. Each indicator function $F_\mu$ should distinguish between metabolites involved only in internal reactions ($\mu\in\mathcal{I}$ for brevity) and metabolites that are exchanged with the surrounding. A convenient parameterisation is given by:
\beeq{
 F_{\mu}\left(y\right)=\left\{
\begin{split}
&\delta\left(y\right)\,,\quad\quad\quad\quad\quad\quad\quad~\quad\,\, \mu\in\mathcal{I}\\
&\int^{M^{\mu}}_{m^\mu} \rho_\mu(z)\delta\left(y-z\right)dz\,,\quad \mu\not\in\mathcal{I}
\end{split}\right.
}
where $\delta(y)$ is a Dirac $\delta$-distribution and $\rho_\mu$ is an {\it a priori} distribution for the exchange rate $\gamma^\mu$. Under this choice, for intracellular metabolites, $F_\mu$ simply enforces the mass-balance constraint (\ref{eq:massbalan}) with $\gamma^\mu=0$ in Equation~(\ref{eq:pdfs}) through a $\delta$-function. For exchanged chemical species, the situation is slightly more complex. If the exchange rate takes a fixed value $z_0$, then $\rho_\mu(z)=\delta(z-z_0)$ and:
\beeq{
 F_{\mu}\left(y\right)=\delta\left(y-z_0\right)
}
corresponding, once inserted in Equation~(\ref{eq:pdfs}), to the mass-balance constraint (\ref{eq:massbalan}) with $\gamma^\mu=z_0$. If, however, the exchanged rate is known only probabilistically, then $\rho_\mu(z)$ can be a non-trivial distribution and $F_\mu$ enforces (\ref{eq:massbalan}) in (\ref{eq:pdfs}) by weighing all possible values of $\gamma^\mu$ according to the measure $\rho_\mu$. For instance, if there is no a\textit{ priori} information about the exchange rate, then $\rho_\mu(z)$ can be taken to be uniform. Note that when $\gamma^\mu$ is a random variable, one can consider it as another unknown rate, so that one could also be interested in estimating its {\it a posteriori} distribution $P_\mu(\gamma)$.
%\footnote{From a point of view of statistical mechanics this is nothing more than estimating the distribution of the local field $\gamma^\mu=\sum_{i=1}^N\xi_i^\mu x^i$.}
The problem we want to face is that of computing quantities like Equation~(\ref{eq:pdfs}) for all $i$'s.

\subsection{Weighted Belief Propagation}\label{ssec:wBP}

To push mathematically forward expression \eqref{eq:pdfs}, we need to do some type of approximation for $S$. Following \cite{Font2012}, we assume the bipartite graph that describes the interdependency of reactions and metabolites to be locally tree-like. In such a case, we are supposing there are no (or only very long) cycles connecting the reactions that process a given metabolite $\mu$. (In the following, we shall write $i \in \mu$ to indicate that reaction $i$ processes, either as a substrate or as a product, metabolite $\mu$.) Thus, if we imagine removing metabolite $\mu$ from the system, all reactions $i\in\mu$ become (approximately) statistically independent, as they belong to separate branches of the metabolic (tree-like) network, and their joint PDF factorizes. This is explained pictorially in Figure \ref{fig1}a. If we now put metabolite $\mu$ back, we see that, at a fixed value, $x$, of reaction $i$, the probability $L_{\mu\to i}(x)$ that mass balance holds for metabolite $\mu$ can be expressed, in terms of the factorized PDF computed in absence of $\mu$, as (see Figure \ref{fig1}b):
\beeq{
L_{\mu\to i}(x)=\frac{1}{L_{\mu\to i}}\int_{D_{\mu\backslash i}} d x_{\mu\backslash i}F_{\mu}\left(\sum_{\ell\in\mu\backslash i}\xi^\mu_\ell x^\ell+\xi^\mu_i x\right)\prod_{\ell\in\mu\backslash i} P_{\ell\to\mu}(x^\ell)\label{ca1}~~
}
with $L_{\mu\to i}$, a normalisation constant. In this formula, we use Latin labels ($i,\ell,\ldots$) for reactions and Greek ones ($\mu,\nu,\ldots$) for metabolites, while the script, $\ell\in\mu\backslash i$, denotes the reactions that process $\mu$ except reaction $i$. Accordingly, we defined the shorthands, $dx_{\mu\backslash i}=\prod_{\ell\in \mu\backslash i} dx^\ell$ and $D_{\mu\backslash i}=\times_{\ell\in \mu\backslash i}[m^{\ell},M^{\ell}]$. The quantity, $P_{\ell\to\mu}(x^\ell)$, is the PDF of flux $\ell$ taking a value $x^\ell$, when metabolite $\mu$ is removed. Those PDFs are, in turn, given by the probability, for each reaction $i$, to satisfy the mass balance conditions for all the metabolites they process, except $\mu$ (see Figure \ref{fig1}c), namely:
\beeq{
P_{i\to\mu}(x)=\frac{1}{P_{i\to\mu}}\prod_{\nu\in i\backslash\mu}L_{\nu\to i}(x)\label{ca2}\,
}
Here, the set $\nu\in i\backslash\mu$ stands for all metabolites $\nu$, processed by reaction $i$, except $\mu$, and $P_{i\to\mu}$ is a normalisation constant. Again, the above equations simply state the fact that, on locally tree-like graphs, the contributions to the PDFs coming from each node (reaction or metabolite) nicely factorize.

\begin{figure}[H]
\centering
\begin{picture}(500,175)

\put(85,165){{(\textbf{a})}}
\put(100,71){$\mu$}
\put(40,65){$\ell \in \mu$}
\put(96,25){$P_{ i\to \mu}(x)$}
\put(72,15){$i$}
\put(20,10){\includegraphics[width=0.28\textwidth, height=0.28\textwidth]{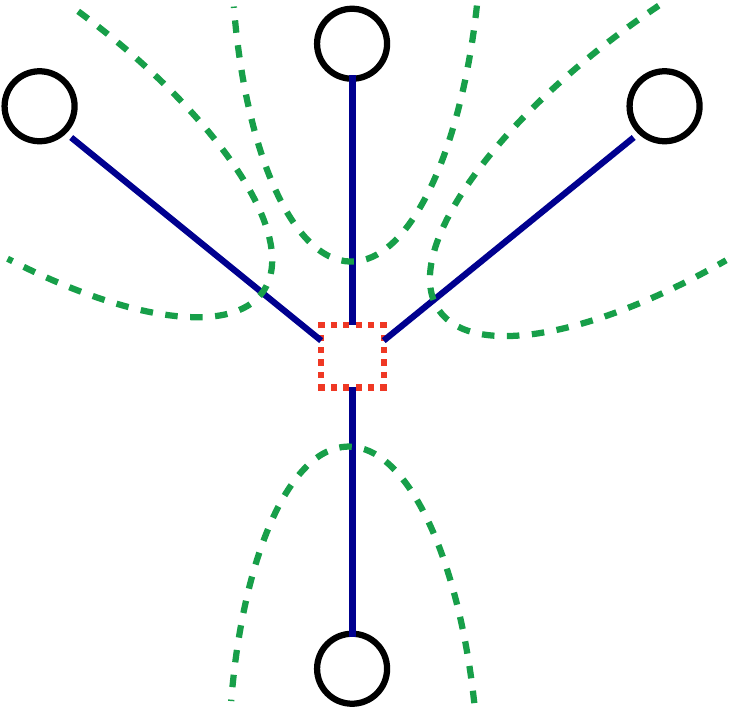}}

\put(260,165){{(\textbf{b})}}
\put(275,15){$x^i=x$}
\put(200,138){$x^\ell$}
\put(275,65){$F_\mu$}
\put(190,92){$P_{ \ell\to \mu}(x^\ell)$}
\put(220,42){$L_{\mu\to i}(x)$}
\put(195,10){\includegraphics[width=0.28\textwidth, height=0.28\textwidth]{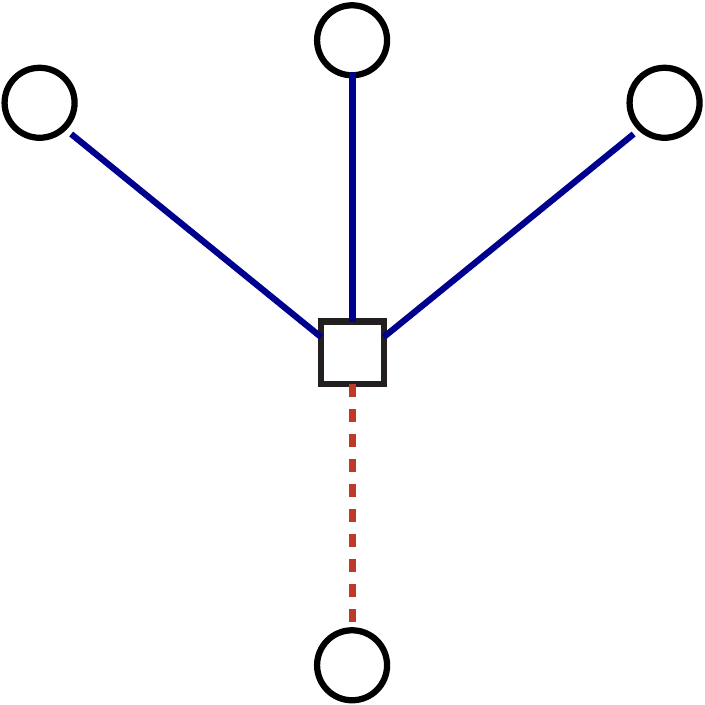}}

\put(425,165){{(\textbf{c})}}
\put(412,140){$\mu$}
\put(442,76){$i$}
\put(452,100){$\eta \in i$}
\put(388,36){$L_{\nu\to i}(x)$}
\put(440,16){$\nu$}
\put(360,10){\includegraphics[width=0.28\textwidth, height=0.28\textwidth]{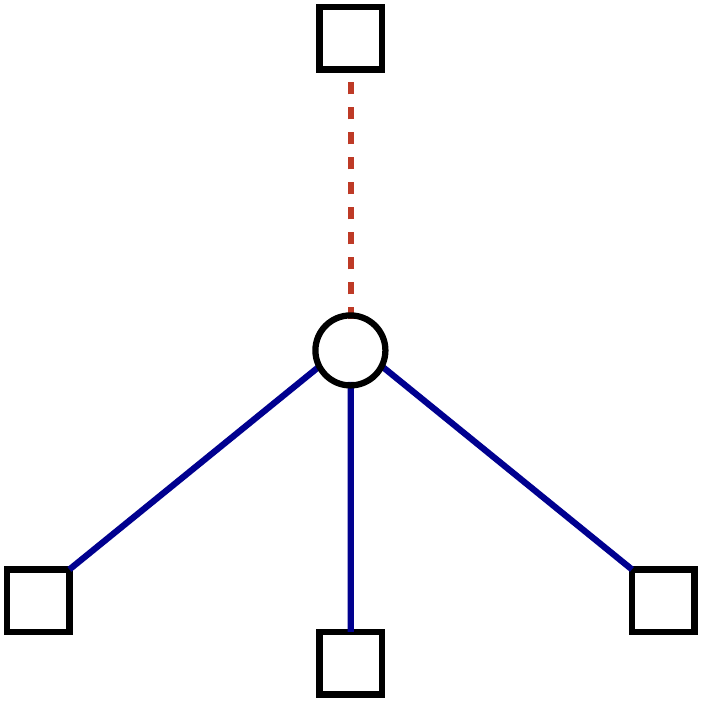}}
\end{picture}
\caption{The method used to derive self-consistency Equations (\ref{ca1}) and (\ref{ca2}) for the conditional marginals; we only show the nearest neighbours of what one must imagine to be a large tree-like bipartite graph, where circles are reactions and squares, metabolites. {(\textbf{a})}~Metabolite $\mu$ and the reactions that process it ($\ell \in \mu$). If we assume removing $\mu$ from the system, all reactions connected to it belong to disjoint branches of the metabolic network, highlighted with the dashed lines. As a consequence, their joint probability distribution function (PDF) factorizes in the product of the marginals, $P_{ \ell\to \mu}(x)$, of each reaction $\ell$. {(\textbf{b})} When metabolite $\mu$ is put back in the graph, the probability, $L_{\mu\to i}(x)$, of satisfying its mass balance condition when fixing the flux of reaction $i$ to $x$ depends on the marginals, $P_{ \ell\to \mu}(x)$, of all neighbours, but $i$, and on the indicator function, $F_\mu$. {(\textbf{c})} The marginal $P_{ i\to \mu}(x)$, which is computed in absence of $\mu$, expresses the probability that $i$ satisfies the mass balance conditions for all the metabolites it processes ($\eta \in i$), except $\mu$. On a tree, each mass balance condition is independent, so that the probability of satisfying all of them is given by the product of the various $L_{\nu\to i}(x)$.}
\label{fig1}
\end{figure}

The conceptual step of removing metabolites from the system is the key that allows us to recast the problem in the set of self-consistency equations \eqref{ca1} and \eqref{ca2}, for the conditional probabilities (the reader should keep in mind that this is, however, just a mathematical trick with no biological interpretation whatsoever \cite{Mezard2009}). Once the fixed point of the system formed by Equations \eqref{ca1} and \eqref{ca2} is known, one can compute the actual PDFs of the fluxes in the metabolic network as:
\beeq{
P_{i}(x)=\frac{1}{P_{i}}\prod_{\nu\in i}L_{\nu\to i}(x)\,
\label{marginal1}
}
where $P_i$ is a normalisation constant. Note that Equation~\eqref{marginal1} also provides the recipe to evaluate the PDFs, $P_\mu(\gamma)$, for the exchange rates, $\gamma^\mu$, once the conditional marginals, $P_{\ell\to\mu}(x^\ell)$, are known.

As discussed in \cite{Font2012,Braunstein2008}, the difficulties of the problem lie not so much in the derivation of Equations \eqref{ca1} and \eqref{ca2}, but in devising an efficient method to solve them. In wBP, we tackle the issue by representing the marginals \eqref{ca2} through a collection of $\mathcal{N}$ variables and associated weights, rather than discretizing them as one would normally do when facing a similar problem. Let us illustrate the idea with a fairly simple example. Consider the integral:
\beeq{
\phi_x(x)=\frac{1}{C}\int_0^1dy\phi_y(y)\int_0^1dz\phi_z(z)\delta(x+y+z-1)\,
\label{example}
}
with the extra condition that $x\geq 0$, where $\phi_y$, $\phi_z$ are known densities normalised in the interval $[0,1]$, and $C$ is a normalisation constant. To evaluate \eqref{example}, we could use Monte Carlo integration and draw $\mathcal{N}$ pairs of random variables $\{(y_i,z_i)\}_{i=1}^{\mathcal{N}}$ according to the distributions, $\phi_y$ and $\phi_z$. Correspondingly, an estimate for $\phi_x(x)$ can be written as:
\beeq{
\phi_x(x)=\frac{\sum_{i=1}^{\mathcal{N}}\delta(x+y_i+z_i-1)\Theta(1-y_i-z_i)}{\sum_{i=1}^{\mathcal{N}}\Theta(1-y_i-z_i)}
}
where the term, $\Theta(1-y_i-z_i)$, accounts for draws for which the quantity $x= 1-y_i-z_i$ must be rejected due to the condition $x\geq 0$. The latter condition indeed defines a feasible triangular region in the integration plane, $yz$ (see Figure \ref{fig2}), such that every extraction $(y_i,z_i)$ falling outside this domain must be rejected. This method (basically, naive Monte Carlo integration) is, hence, poised to be rather inefficient. Fortunately, we know precisely where the rejection region is, and we can rewrite Equation~\eqref{example} as follows:
 \beeq{
\phi_x(x)=\frac{1}{C}\int_0^1dy\phi_y(y)\int_{0}^{1-y}dz\phi_z(z)\delta(x+y+z-1)\,
\label{example2}
}
Now, Equation~\eqref{example2} does not contain a rejection region, but we cannot apply Monte Carlo integration just yet, since $\phi_z(z)$ is not normalised in the interval $[0,1-y]$. Introducing the corresponding weight:
 \beeq{
w(y)\equiv\int_{0}^{1-y}dz\phi_z(z)
}
we can, however, re-cast Equation~\eqref{example2} in the form:
 \beeq{
\phi_x(x)=\frac{1}{C}\int_0^1dy\phi_y(y)\int_{0}^{1-y}dz\phi_z(z|y) w(y)\delta(x+y+z-1)\
\label{example3}
}
with $\phi_z(z|y)\equiv\phi(z)/w(y)$. The distributions appearing above are now properly normalized. Therefore, to evaluate Equation~\eqref{example3}, we can simply draw $\mathcal{N}$ pairs $\{(y_i,z_i)\}_{i=1}^{\mathcal{N}}$ according to $\phi_y(y)$ and $\phi_z(z|y)$, respectively, and estimate $\phi_x(x)$ by:
\beeq{
\phi_x(x)=\sum_{i=1}^{\mathcal{N}} \alpha_i\delta(x+y_i+z_i-1)\,,\quad\quad \text{with }~\alpha_i\equiv \frac{w(y_i)}{\sum_{j=1}^{\mathcal{N}} w(y_j)}
}
\emph{i.e}., by $\mathcal{N}$ pairs of variables and weights $\{(x_i\equiv 1-y_i-z_i,~\alpha_i)\}_{i=1}^\mathcal{N}$.

The key point of this method is that the reweighted density, $\phi_z(z|y)$, has a $y$-dependent support, such that rejection never occurs. Thus, at a price of computing a weight, $w(y)$, we overcome the whole rejection issue, and the method becomes much more efficient.

The great advantage of using wBP is that, at fixed $\mathcal{N}$, its running time goes as $\mathcal{O}(2 N k)$, where $k$ is the average number of metabolites processed by each reaction. Thus, as opposed to sampling techniques that have normally super-linear mixing times \cite{Simonovits2003}, wBP only scales linearly with the number of reactions (see Figure \ref{fig:BPtime}), making it an ideal candidate for application to genome-scale metabolic networks. In the present work, we focus, however, on the relatively small case of the hRBC, so that we are able to compare with sampling methods that yield a uniform exploration of the solution space $S$ \cite{Lovasz1999}. Due to the nature of such methods (see next section), this type of comparison is still not feasible for larger systems. This, and the fact that previous results are available \cite{Braunstein2008}, make the metabolic network of the hRBC the ideal testing ground for wBP.

\begin{figure}[H]
\begin{picture}(300,175)
\put(230,110){$\text{{\small Rejection Region}}$}
\put(250,90){ ($x\leq 0$)}
\put(195,145){$z$}
\put(335,0){$y$}
\put(200,10){\includegraphics[width=5cm, height=5cm]{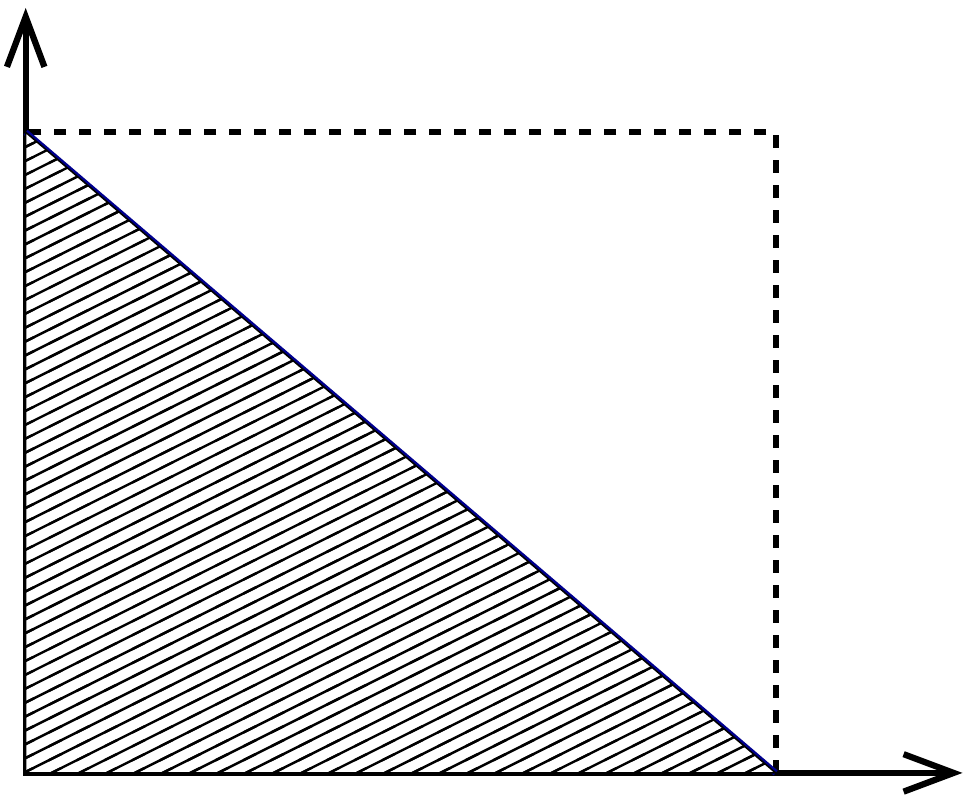}}
\end{picture}
\caption{Avoiding rejection. As explained in the text, from the original integration region, $(y,z)\in[0,1]\times [0,1]$, only the one below the line $z=1-y$ contributes to $\phi_x(x)$. However, in this lower triangle, the density, $\phi_y(y)\phi_z(z)$, is no longer normalised. This is easily dealt with by reweighting the integral.}
\label{fig2}
\end{figure}
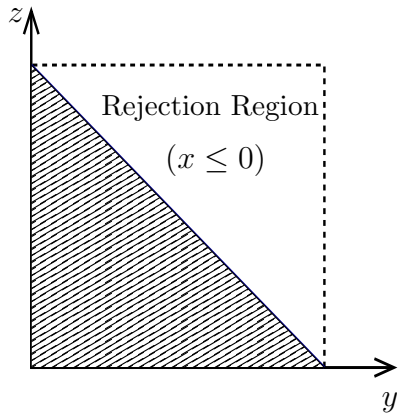

\begin{figure}[H]
\centering
\includegraphics[width=0.6\textwidth]{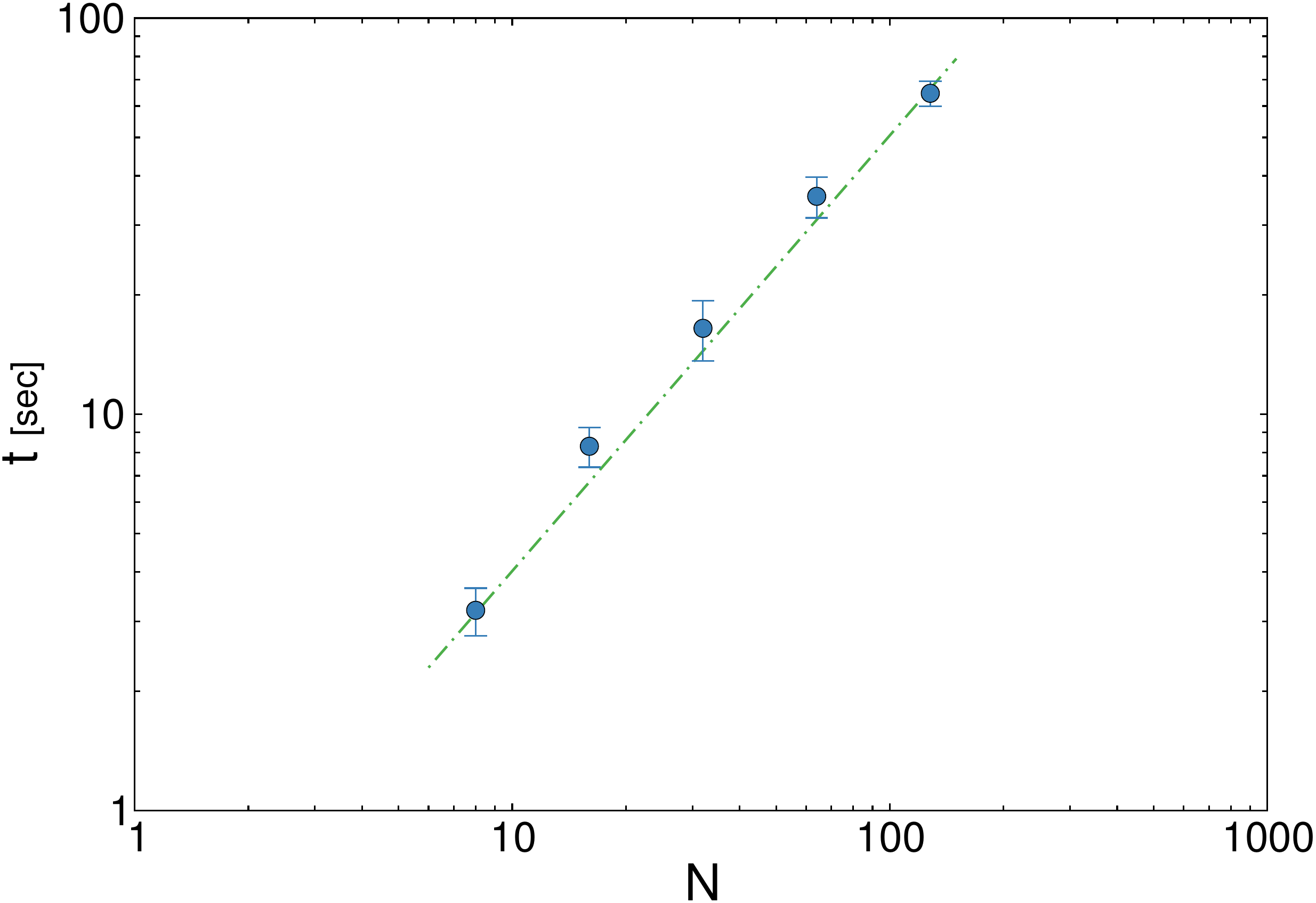}
\caption{The running time $t$ of the weighted Belief Propagation (wBP) algorithm \emph{vs}. the number of reactions, $N$. For each value of $N$, we average here over 10 random synthetic metabolic networks, each having $M=N/2$ metabolites. The algorithm (blue circles) scales linearly with the system size; a linear function, $t \propto N$ (green dashed line), is plotted to guide the eye.}
\label{fig:BPtime}
\end{figure}

\subsection{The Kernel Hit-and-Run (KHR) Algorithm}

In order to sample the solution space $S$ and obtain exact PDFs of individual fluxes for the hRBC by a controlled method that guarantees uniformity, we have developed an optimized version of Hit-and-Run Monte Carlo, which we call the {\it Kernel Hit-and-Run} method. Let us start by re-writing constraints (\ref{eq:massbalance}) explicitly for metabolites involved in internal reactions and the rest:
\beaeq{
\sum_{i\in \mu}\xi_i^\mu x^i=0\,,&&\quad \mu\in\mathcal{I}\label{internal}\\
m^\mu\leq\sum_{i\in \mu}\xi_{i}^\mu x^i\leq M^\mu\,,&&\quad \mu\not\in \mathcal{I}\label{rest}\\
m^i\leq x^i\leq M^i\,,&&\quad i=1,\ldots, N\label{bounds}
}
We note that the set of $|\mathcal{I}|$ equations in \eqref{internal} defines the null-space of $\bm{\xi}$, and geometrically corresponds to a family of hyperplanes passing through the origin $\bm{x=0}$. Let us denote the dimension of the null space of $\bm{\xi}$ as $K$. Clearly, $K$ would be at least $N-|\mathcal{I}|$ (actually $K=N-|\mathcal{I}|$ when $\bm{\xi}$ has full row rank, which can always be made to be the case and which we assume from now on). This means, obviously, that, although the number of variables in the system is $N$, due to the constraints in the model, the actual dimension of the solution space $S$ is only $K$. As in real metabolic networks most reactions are internal, the dimension $K$ of the null space will be significantly smaller than the original dimension of the problem $N$. Additionally, it turns out that the way to implement in practice such a dimensional reduction is quite straightforward: suppose that a basis of the null-space has been found, e.g., through Gaussian elimination or singular value decomposition (SVD), and let us denote as $\bm{y}=(y^1,\ldots,y^K)$ the system of coordinates with respect to such a basis, so that we can write each flux in this basis as $x^{i}=\sum_{j=1}^{K}\Phi^{i}_{j}y^j$, with $\bm{\Phi}$ an $N\times K$ matrix related to the change of basis between the original space and the null subspace. Plugging this into Equations~\eqref{rest} and \eqref{bounds} allows us to write:
\beeq{
m^\mu\leq\sum_{j=1}^{K} \Psi^{\mu}_j y^j\leq M^\mu\,,&\quad\quad\quad \mu\not\in \mathcal{I}\\
m^i\leq \sum_{j=1}^{K}\Phi^{i}_{j}y^j\leq M^i\,,&\quad\quad\quad i=1,\ldots, N
\label{eq:projected polytope}
}
where we have defined the projected stoichiometric matrix, $\Psi$, with entries $\Psi^{\mu}_j=\sum_{i=1}^N\xi_{i}^\mu \Phi^{i}_{j}$. The set of Equations \eqref{eq:projected polytope} defines a $K$-dimensional polytope in the null space (see Figure \ref{fig3}), which can be sampled {\it uniformly} by using the Hit-and-Run algorithm \cite{Lovasz1999, Smith1984,Berbee1987}. Finally, to go back to the original space, that of the reaction rates, we simply use the fact that $x^{i}=\sum_{j=1}^{K}\Phi^{i}_{j}y^j$. The sampling properties of the \textit{Hit-and-Run algorithm} under the uniform measure were indeed mathematically proven \cite{Lovasz1999}, and in our case, it is very easy to see that the uniform measure in the $K$-dimensional null space is preserved under a linear transformation, so that the final sample in the full-dimensional space is also uniform by~construction.

While the sampling measure of KHR is well controlled, a word needs to be spent on the algorithmic mixing time. For the standard {\em Hit-and-Run} algorithm, this scales as the square of the dimensions times the diameter of the polytope, \emph{i.e}., in practice, cubically with the number of dimensions \cite{Lovasz1999, Simonovits2003}. Yet, as mentioned before, in our approach, the dimension of the polytope is $K$, rather than $N$. This can yield a significant reduction in computation times if $K$ is small compared to $N$, as will quite generally be the case. For the hRBC, for instance, we pass from $N=46$ (which can be problematic, e.g., for Monte Carlo rejection \cite{Wiback2004}), to a much more modest $K=12$, which is sampled quite fastly by KHR. Note also that no additional constraints need to be introduced to enclose the polytope (as opposed to \cite{Almaas2004}). This is due to the fact that the $M-|\mathcal{I}|$ metabolites that are exchanged with the environment suffice to bound the polytope in the null space.

\begin{figure}[H]
\begin{picture}(300,230)
\put(130,155){$y_{k}$}
\put(280,40){$y_{j}$}
\put(355,165){Polytope}
\put(190,220){$K$-dimensional null space}
\put(150,10){\includegraphics[width=9cm, height=7cm]{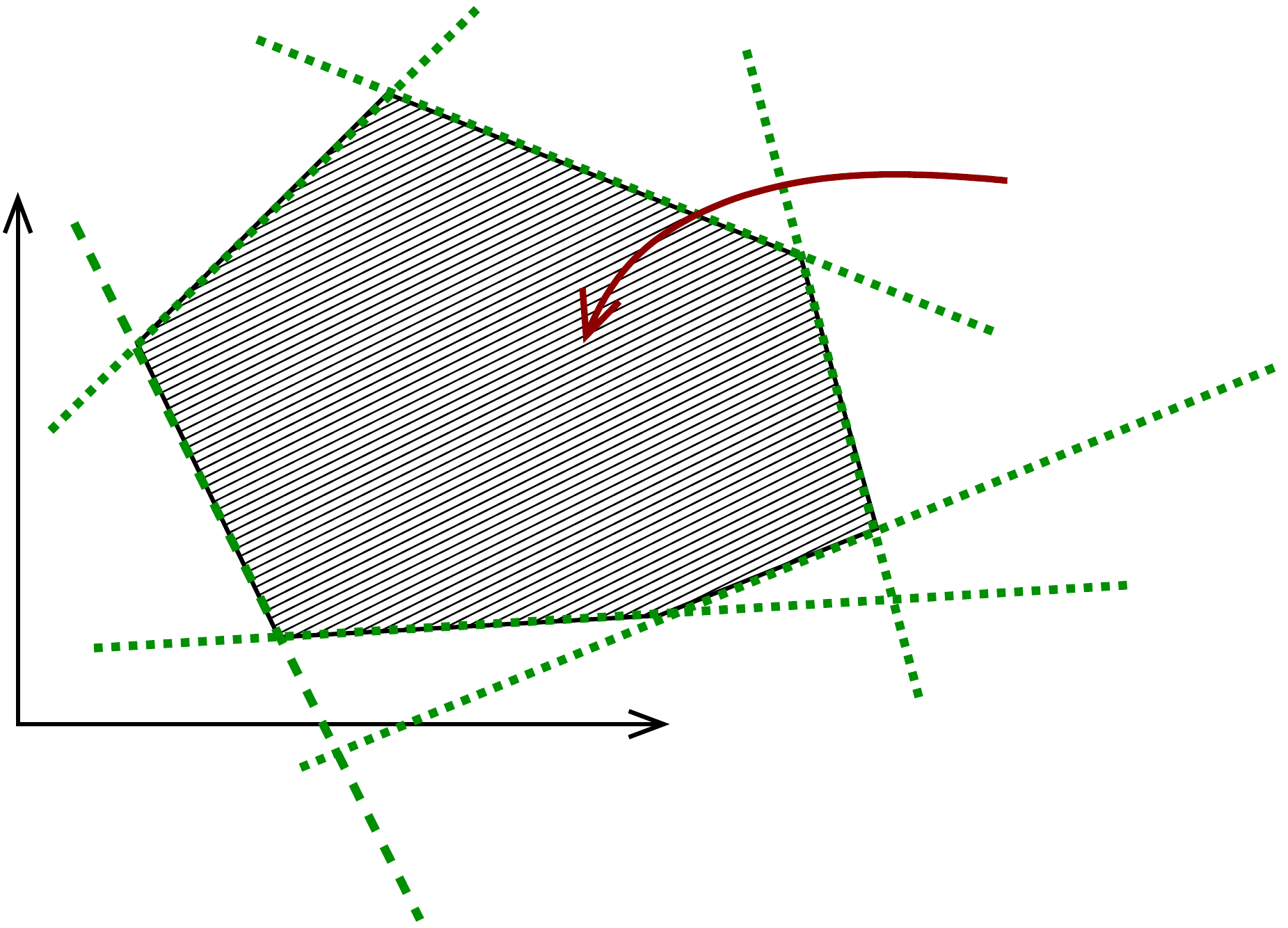}}
\end{picture}
\caption{A cartoonish representation of the polytope in a $K$-dimensional null space spanned by $y$-coordinates. Here, the green dashed lines represent the set of hyperplanes \eqref{eq:projected polytope} enclosing the polytope.}
\label{fig3}
\end{figure}
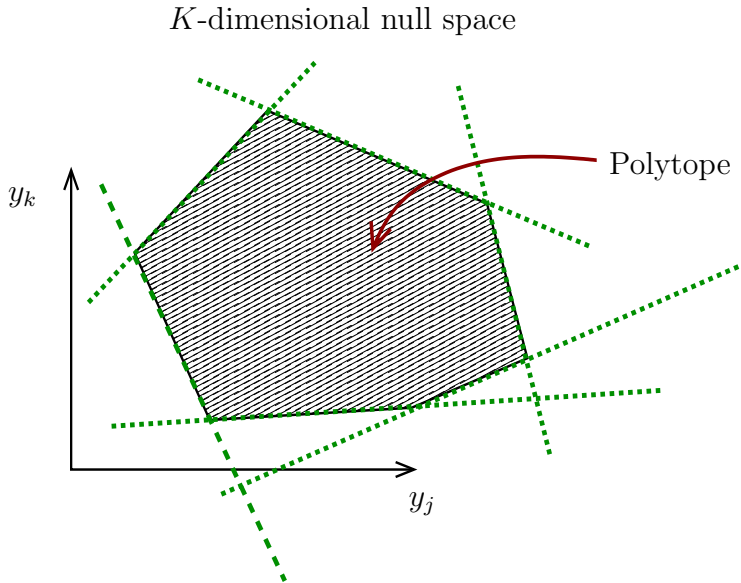

Finally, it is worth mentioning that the matrix $\bm{\Phi}$ can be easily obtained with any standard algebra software or by standard SVD%please define -- defined 2 paragraphs above
~algorithms, and that no matrix inversion is required to compute the projected matrix $\Psi$ nor to convert the obtained sample to the original, full dimensional space. Therefore, to sum up, KHR uniformly samples the solution space $S$, with a mixing time that scales as $\mathcal{O}(K^3$).

\section{Results and Discussion}

We have applied the wBP and KHR algorithms to the study of the metabolic network of the hRBC. As mentioned in Section \ref{ssec:wBP}, such a choice is dictated by the fact that the hRBC size allows us to apply HR in a modest time and that previous results are available. We have used the same network considered in \cite{Braunstein2008,Wiback2004,Price2004}, which accounts for 34 metabolites, 32 internal reactions and 14 exchange reactions (see Figure \ref{fig4}). We are able to smoothly apply the method by using the effective reaction domains computed in \cite{Wiback2004} and used also in \cite{Braunstein2008}. Such domains are derived starting from real enzymatic rates \cite{Wiback2003, Wiback2002}, so that they are physiologically meaningful, and span several orders of magnitude. Note that this would be a major issue if one were to discretize marginals \eqref{ca2}, as it would require dealing with binning functions defined on totally different scales. Thanks to our representation in terms of variables and weights, the fact that the reaction fluxes are of a different order of magnitude does not affect our method at all.

We run the wBP algorithm by representing the marginals, like Equation~\eqref{ca2}, with sets of variables/weights $\{(x_i,~\alpha_i)\}_{i=1}^\mathcal{N}$ of size $\mathcal{N}=500$. To solve the fixed point equations \eqref{ca1} and \eqref{ca2}, we performed 30 iterations of our method. We started with uniform weights $\{\alpha_\ell\}_{\ell=1}^\mathcal{N}$ and, at each iteration $t=1,\ldots,30$, and for each fixed value of the variable $x_i$, we applied wBP $10^3\times t$ times to evaluate the average weight $\alpha_i$. Once convergence was reached, we used the variable/weight sets to compute the final 46 PDFs, $P_{i}(x)$ and $P_\mu(\gamma)$, according to Equation \eqref{marginal1}. In this last step, we averaged the weight values over $10^5$ wBP extractions to achieve a higher accuracy. We report the results in Figures \ref{fig4} and \ref{fig5}, where we compare our method with KHR; the agreement is excellent. The reaction PDFs obtained with both methods have indeed a very similar domain and shape in most of the cases. Notably, wBP does not perfectly capture the profile of reactions involving currency metabolites, such as ATP, ADP, NADP and NADPH%please define abbreviations -- Is this realy necessary? I think these molecules are known more after their abbreviation than their full names, which is considerably long and would unnecessarily increase the text length
. An explanation of this may lie in the fact that these compounds are highly connected in metabolic networks and likely to be involved in small loops that are not considered by the wBP method.
\begin{figure}[H]
\centering
\includegraphics[width=14cm, height=12.25cm]{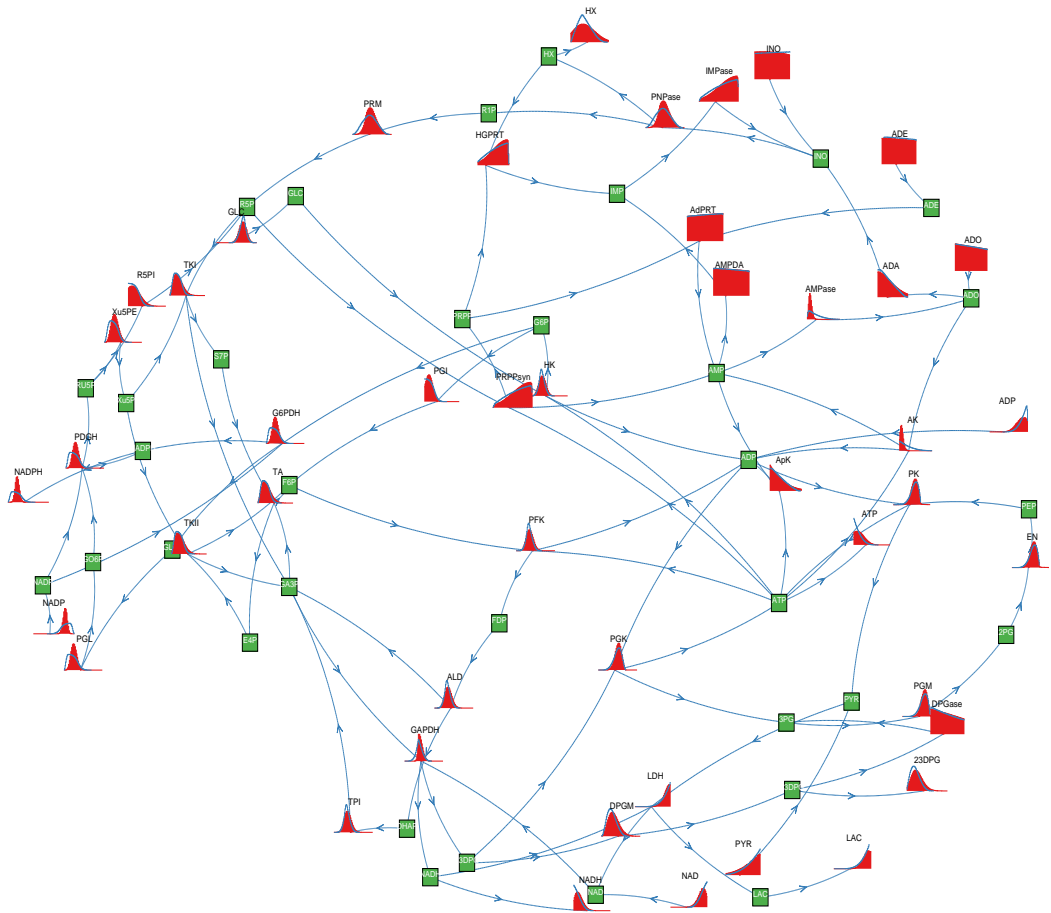}
\caption{Results for human red blood cell. Here, we draw a pictorial representation of the system as a directed bipartite graph. Reaction nodes are plotted with their PDFs and metabolite nodes with green squares. Arrows entering (resp. leaving) a reaction stand for a substrate (resp. a product). We have plotted the marginals, $P_i (x)$, for the internal reactions together with the $P_\mu (\gamma)$ for the exchange rates (these are the leaves on the bipartite graph). For the densities, we have used the wBP method (red filled plots) and have compared them with the KHR algorithm (blue solid lines).}
\label{fig4}
\end{figure}

\begin{figure}[H]
\centering
\includegraphics[width=16cm]{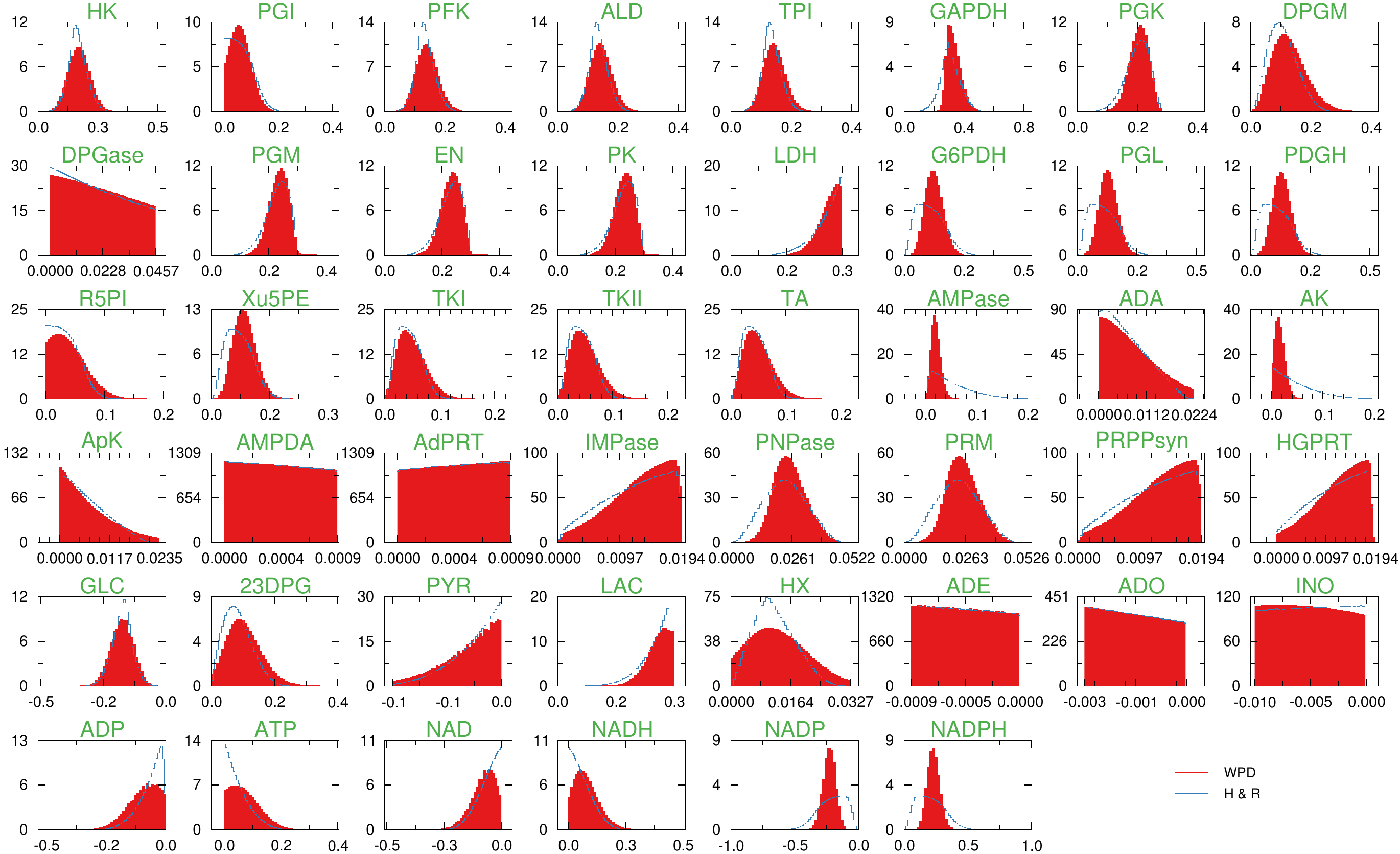}
\caption{Results for human red blood cell. The probability density functions of the reaction rates; reaction names are the same as \cite{Braunstein2008}. For the densities, we have used the wBP method (red filled plots) and have compared them with the Kernel Hit-and-Run (KHR) algorithm (blue solid lines). Note that the flux ranges span different orders of magnitude, but still, the profiles are very smooth for both weighted population dynamics and the Kernel Hit-and-Run algorithm.}
\label{fig5}
\end{figure}

Concerning the results obtained by KHR, we have been particularly careful to make sure we obtain a uniform distribution of the solution space $S$. As the uniform measure is guaranteed to be a limiting distribution, the difficulty lies, of course, in deciding when such a limit has practically been reached in the simulations. In this regard, we decided to apply three conservative measures to ensure this: first of all, we checked that the initial conditions did not affect the results and, after that, the simulations were averaged over the initial conditions ({the initial points were generated using the MinOver algorithm~\cite{Krauth1987}); secondly, the initial time-window was disregarded in the sampling; and finally, in order to avoid correlation effects in the sampling of the PDFs, we only recorded points at large spacing in time.
In Figure \ref{fig5}, we report a panel with all the PDFs for the hRBC network to make the comparison between the methods more straightforward.

%%%%%%%%%%%%%%%%%%%%%%%%%%%%%%%%%%%%%%%%%%

\section{Conclusions}

In this work, inspired by techniques employed in the statistical mechanics of disordered systems, we have presented a novel method to estimate distributions of reaction fluxes in constraint-based models of metabolic networks. The wBP methodology has, in our view, clear advantages when compared with alternative approaches. If compared to rejection-based Monte Carlo methods \cite{Price2004} or even to more refined sampling approaches \cite{Almaas2004}, our algorithm has the significant advantage of scaling linearly with the system size, a feature that makes it particularly suitable to study genome-scale metabolic reconstructions. Comparing wBP with similar MP-based approaches \cite{Braunstein2008}, our method turns out to be unaffected by the flux ranges spanning several orders of magnitude nor by the stoichiometric coefficients taking real values. Note that the former property is particularly important for the study of real metabolic systems in physiologic conditions, as enzymatic rates can vary wildly across the network \cite{Wiback2004,Price2004}.

wBP can also be integrated with optimization-based flux balance analysis (FBA), as it easily allows us to evaluate the PDFs of the enzymatic rates close to optimality (assuming a score function is known) by just injecting \textit{a priori} to keep fluxes ``close'' to their optimal value.

We have also compared the performance of wBP against the KHR method. The latter is a controlled Hit-and-Run Monte Carlo taking place in the null space defined by the set of internal reactions, where a considerable effective dimensional reduction can be achieved. Indeed, starting from the original $N$-dimensional space of solutions, one can wind up into a space that, in the best of cases, has the same number of dimensions as the number of exchange reactions. Given the considerable gains that this observation provides, we believe that this method may be worth being explored further in its own right.

The validation of the wBP algorithm in the hRBC network, which can be considered a benchmark for the sampling problem for constrained metabolic models, opens the door to future applications of the method to more relevant organisms, such as {\em Mycoplasma pneumoniae} \cite{Wodke2013} or {\em Escherichia coli} \cite{Feist2007}. In our opinion, the fact that its algorithmic complexity scales linearly with the number of reactions, combined with the aforementioned ability to deal with more realistic bounds, makes it a highly promising candidate for effective solutions to this challenging task.

%%%%%%%%%%%%%%%%%%%%%%%%%%%%%%%%%%%%%%%%%%

\acknowledgements{Acknowledgments}

We thank de Martino, D., G\"uell, O., Guimer\`a, R., Sales-Pardo, M., Serrano, M.A. and Sagu\'es,~F. for useful discussions and comments. F.F.C. would like to thank funding from MINECO%please define
 (grant FIS2012-31324) and AGAUR%please define
 (grant 2012FI\_B00422). F.A.M. acknowledges financial support from European Union Grants, PIRG-GA-2010-277166 and PIRG-GA-2010-268342. A.D.M. is supported by the DREAM Seed Project of the Italian Institute of Technology (IIT). The IIT Platform Computation is gratefully acknowledged.

%%%%%%%%%%%%%%%%%%%%%%%%%%%%%%%%%%%%%%%%%%

\conflictofinterests{Conflicts of Interest}
The authors declare no conflict of interest.

%=================================================================
% References: Variant B
%=================================================================
% Use the following option to include external BibTeX files:
%\bibliography{lite}

\begin{thebibliography}{----}
\providecommand{\natexlab}[1]{#1}

\bibitem[Bowman \em{et~al.}(1997)Bowman, Churcher, Badcock, Brown,
 Chillingworth, et~al.]{Bowman1997}
Bowman, S.; Churcher, C.; Badcock, K.; Brown, D.; Chillingworth, T.; Connor, R.; Dedman,~K.; Devlin, K.; Gentles, S.; Hamlin, N.
\newblock The nucleotide sequence of Saccharomyces cerevisiae chromosome XIII.
\newblock {\em Nature} {\bf 1997},
\newblock {\em 387},~90--92.

\bibitem[Feist \em{et~al.}(2007)Feist, Henry, Reed, Krummenacker, Joyce,
 et~al.]{Feist2007}
Feist, A.; Henry, C.; Reed, J.; Krummenacker, M.; Joyce, A.; Karp, P.D.; Broadbelt, L.J.; Hatzimanikatis, V.; Palsson, B.\O.
\newblock A genome-scale metabolic reconstruction for {\em {Escherichia coli}}
 K-12 MG1655 that accounts for 1260 ORFs and thermodynamic information.
\newblock {\em Mol. Syst. Biol.} {\bf 2007}, doi:10.1038/msb4100155.

\bibitem[Thiele and Palsson(2010)]{Thiele2010}
Thiele, I.; Palsson, B.O.
\newblock A protocol for generating a high-quality genome-scale metabolic
 reconstruction.
\newblock {\em Nat. Protoc.} {\bf 2010},
\newblock {\em 5},~93--121.

\bibitem[Thiele \em{et~al.}(2013)Thiele, Swainston, Fleming, Hoppe, Sahoo,
 Aurich, Haraldsdottir, Mo, Rolfsson, Stobbe, Thorleifsson, Agren, Bolling,
 Bordel, Chavali, Dobson, Dunn, Endler, Hala, Hucka, Hull, Jameson, Jamshidi,
 Jonsson, Juty, Keating, Nookaew, Le~Novere, Malys, Mazein, Papin, Price,
 Selkov~Sr, Sigurdsson, Simeonidis, Sonnenschein, Smallbone, Sorokin, van
 Beek, Weichart, Goryanin, Nielsen, Westerhoff, Kell, Mendes, and
 Palsson]{Thiele2013}
Thiele, I.; Swainston, N.; Fleming, R.M.T.; Hoppe, A.; Sahoo, S.; Aurich, M.K.;
 Haraldsdottir, H.; Mo, M.L.; Rolfsson, O.; Stobbe, M.D.; \emph{et al}.
\newblock A community-driven global reconstruction of human metabolism.
\newblock {\em Nat. Biotechnol.} {\bf 2013},
\newblock {\em 31},~419--425.

\bibitem[Kauffman \em{et~al.}(2003)Kauffman, Prakash, and
 Edwards]{Kauffman2003}
Kauffman, K.J.; Prakash, P.; Edwards, J.S.
\newblock {Advances in flux balance analysis}.
\newblock {\em Curr. Opin. Biotechnol.} {\bf 2003},
\newblock {\em 14},~491--496.

\bibitem[Orth \em{et~al.}(2010)Orth, Thiele, and Palsson]{Orth2010}
Orth, J.; Thiele, I.; Palsson, B.
\newblock What is flux balance analysis?
\newblock {\em Nat. Biotechnol.} {\bf 2010},
\newblock {\em 28},~245--248.

\bibitem[Schellenberger and Palsson(2009)]{Schellenberger2009}
Schellenberger, J.; Palsson, B.{\O}.
\newblock Use of randomized sampling for analysis of metabolic networks.
\newblock {\em J. Biol. Chem.} {\bf 2009},
\newblock {\em 284},~5457--5461.

\bibitem[Lov{\'a}sz(1999)]{Lovasz1999}
Lov{\'a}sz, L.
\newblock Hit-and-run mixes fast.
\newblock {\em Math. Progr.} {\bf 1999},
\newblock {\em 86},~443--461.

\bibitem[Braunstein \em{et~al.}(2008)Braunstein, Mulet, and
 Pagnani]{Braunstein2008}
Braunstein, A.; Mulet, R.; Pagnani, A.
\newblock Estimating the size of the solution space of metabolic networks.
\newblock {\em BMC Bioinforma.} {\bf 2008}, doi:10.1186/1471-2105-9-240.

\bibitem[Mezard and Montanari(2009)]{Mezard2009}
Mezard, M.; Montanari, A.
\newblock {\em Information, Physics, and Computation}; Oxford University Press: Oxford, UK,
\newblock 2009.

\bibitem[Font-Clos \em{et~al.}(2012)Font-Clos, Massucci, and {P\'erez
 Castillo}]{Font2012}
Font-Clos, F.; Massucci, F.A.; {P\'erez Castillo}, I.
\newblock A weighted belief-propagation algorithm for estimating volume-related
 properties of random polytopes.
\newblock {\em J. Stat. Mech. Theory Exp.} {\bf
 2012}, doi:10.1088/1742-5468/2012/11/P11003.

\bibitem[Price \em{et~al.}(2004)Price, Schellenberger, and Palsson]{Price2004}
Price, N.D.; Schellenberger, J.; Palsson, B.O.
\newblock Uniform sampling of steady-state flux spaces: Means to design
 experiments and to interpret enzymopathies.
\newblock {\em Biophys. J.} {\bf 2004},
\newblock {\em 87},~2172--2186.

\bibitem[Almaas \em{et~al.}(2004)Almaas, Kovacs, Vicsek, Oltvai, and
 Barabasi]{Almaas2004}
Almaas, K.; Kovacs, B.; Vicsek, T.; Oltvai, Z.M.; Barabasi, A.L.
\newblock Global organization of metabolic fluxes in the bacterium
 \emph{{E}}{\em scherichia coli}.
\newblock {\em Nature} {\bf 2004},
\newblock {\em 427},~839--843.

\bibitem[Simonovits(2003)]{Simonovits2003}
Simonovits, M.
\newblock How to compute the volume in high dimension?
\newblock {\em Math. Progr.} {\bf 2003},
\newblock {\em 97},~337--374.

\bibitem[Smith(1984)]{Smith1984}
Smith, R.L.
\newblock Efficient {Monte Carlo} procedures for generating points uniformly
 distributed over bounded regions.
\newblock {\em Oper. Res.} {\bf 1984}, \emph{32},
\newblock 1296--1308.

\bibitem[Berbee \em{et~al.}(1987)Berbee, Boender, Ran, Scheffer, Smith, and
 Telgen]{Berbee1987}
Berbee, H.; Boender, C.; Ran, A.R.; Scheffer, C.; Smith, R.; Telgen, J.
\newblock Hit-and-run algorithms for the identification of nonredundant linear
 inequalities.
\newblock {\em Math. Progr.} {\bf 1987},
\newblock {\em 37},~184--207.

\bibitem[Wiback \em{et~al.}(2004)Wiback, Famili, Greenberg, and
 Palsson]{Wiback2004}
Wiback, S.J.; Famili, I.; Greenberg, H.J.; Palsson, B.{\O}.
\newblock Monte Carlo sampling can be used to determine the size and shape of
 the steady-state flux space.
\newblock {\em J. Theor. Biol.} {\bf 2004},
\newblock {\em 228},~437--447.

\bibitem[Wiback \em{et~al.}(2003)Wiback, Mahadevan, and Palsson]{Wiback2003}
Wiback, S.J.; Mahadevan, R.; Palsson, B.{\O}.
\newblock Reconstructing metabolic flux vectors from extreme pathways: Defining
 the α-spectrum.
\newblock {\em J. Theor. Biol.} {\bf 2003},
\newblock {\em 224},~313--324.

\bibitem[Wiback and Palsson(2002)]{Wiback2002}
Wiback, S.J.; Palsson, B.O.
\newblock Extreme pathway analysis of human red blood cell metabolism.
\newblock {\em Biophys. J.} {\bf 2002},
\newblock {\em 83},~808--818.

\bibitem[Krauth and Mezard(1987)]{Krauth1987}
Krauth, W.; Mezard, M.
\newblock Learning algorithms with optimal stability in neural networks.
\newblock {\em J. Phys. A} {\bf 1987}, doi:10.1088/0305-4470/20/11/013.
\bibitem[Wodke \em{et~al.}(2013)Wodke, Puchalka, Lluch-Senar, Marcos, Yus,
 Godinho, Gutierrez-Gallego, dos Santos, Serrano, Klipp, and Maier]{Wodke2013}
Wodke, J.A.H.; Puchalka, J.; Lluch-Senar, M.; Marcos, J.; Yus, E.; Godinho, M.;
  Gutierrez-Gallego, R.; dos Santos, V.A.P.M.; Serrano, L.; Klipp, E.; Maier,
 T.
\newblock Dissecting the energy metabolism in Mycoplasma pneumoniae through
 genome-scale metabolic modeling.
\newblock {\em Mol. Syst. Biol.} {\bf 2013},
\newblock {\em 9}, doi:10.1038/msb.2013.6.

\end{thebibliography}
\bibliographystyle{mdpi}

\end{document}